\documentclass{IEEEtaes}

\usepackage{color,array,amsthm}
\usepackage{graphicx}
\usepackage{subcaption}
\usepackage{mathtools}
\usepackage{booktabs}

\jvol{XX}
\jnum{XX}
\jmonth{XXXXX}
\paper{1234567}
\pubyear{2020}
\doiinfo{TAES.2020.Doi Number}

\begin{document}

\pagestyle{empty}
\pagenumbering{gobble}

\title{Contact Plan Design For Optical Interplanetary Communications} 

\author{Jason Gerard}
\affil{Concordia University, Montreal, Quebec, Canada} 

\author{Juan A. Fraire}
\affil{Inria, INSA Lyon, Villeurbanne, France\\CONICET - Universidad Nacional de Córdoba, Argentina}  

\author{Sandra Céspedes}
\affil{Concordia University, Montreal, Quebec, Canada} 

\receiveddate{Manuscript received XXXXX 00, 0000; revised XXXXX 00, 0000; accepted XXXXX 00, 0000.\\
The authors would like to acknowledge the support from Concordia's PhD Fellowship Award, NSERC Canada Discovery Grant RGPIN-2024-05730, ANID Basal Project CIA250006, and the French National Research Agency (ANR) under project ANR-22-CE25-0014-01.}

\corresp{(Corresponding author: J. Gerard email: jason.gerard@mail.concordia.ca)}. 

\authoraddress{J. Gerard and S. Céspedes are with the Department of Computer Science and Software Engineering, Concordia University, Montréal, QC, H2W 2L4, Canada.\\
J. Fraire is with Inria, INSA Lyon, France and CONICET-Universidad Nacional de Córdoba, Córdoba, Argentina}


\maketitle

\begin{abstract}
Space exploration missions generate rapidly increasing volumes of scientific telemetry that far exceed the capacity of today’s manually scheduled, RF-based deep-space infrastructure. 
Free-space optical (FSO) communications promise orders of magnitude higher throughput, but their narrow beams require precise pointing, acquisition, and tracking (PAT) for link establishment and tightly synchronized contact schedules. 
Critically, no existing contact plan design (CPD) framework accounts for optical head retargeting delay, the time spent during coarse pointing and link acquisition before data transmission begins, which directly reduces usable contact time.
Retargeting delay is the dominant impairment unique to optical networks, which induces a seconds-to-minutes-long mechanical pointing process for an optical terminal’s laser from its current partner to the next receiver. 
This paper introduces the first PAT-aware CPD framework for optical interplanetary backhaul networks, combining a temporal network capacity model with a mixed-integer linear programming (MILP) scheduler that embeds a physics-driven retargeting delay model. 
The model captures directional temporal flows across both direct-to-Earth optical links and two-hop relay paths using delay/disruption-tolerant networking (DTN) satellites, providing a comparative evaluation against heuristic and delay-unaware models. 
We also introduce an optical network duty-cycle metric that quantifies the proportion of time spent transmitting to the contact window duration, exposing capacity lost to retargeting delay, an issue not analyzed in prior work. 
Using realistic orbital dynamics and communication models, our results show that our MILP scheduler delivers over 30\% higher network capacity than a greedy algorithm implementing the same underlying flow model. 
More importantly, the results uncover a fundamental behavioral shift: when retargeting delays are modeled accurately, optimal schedules favor fewer but longer optical links that maximize throughput while minimizing retargeting overhead. 
These findings demonstrate that zero-delay assumptions substantially overestimate achievable performance and yield unrealistic contact plans, underscoring the need for PAT-aware temporal modeling in future autonomous optical backhaul networks. 
\end{abstract}


\section{Introduction}

The space industry is experiencing an unprecedented wave of investment and enthusiasm, often referred to as the "New Space" era. 
Breakthroughs, such as reusable rockets and CubeSat-standard satellites, have dramatically lowered the costs of space exploration, making it more accessible than ever before. 
This technological revolution is propelling various ambitious missions, including NASA's Artemis and Mars Exploration Programs, SpaceX’s Starship missions, and international initiatives such as ESA's Moonlight project~\cite{israel_lunanet_2020}.
Space missions often involve rovers, landers, probes, and low-cost wireless sensors, each producing large amounts of data that must be transmitted through deep space links back to Earth for analysis~\cite{ledesma_architectural_2024}.
However, NASA's deep space network (DSN) has a limited bandwidth, operates within constrained radio frequency (RF) communication bands, and requires manual intervention from mission operators to schedule transmissions due to insufficient network autonomy~\cite{edwards_addressing_2024}.
The demand for interplanetary communication systems has outgrown the current capabilities of the DSN.
One example is the Mars Relay Network. 
It has been reported that, at certain times, up to 75\% of mission data could not be transmitted back to Earth due to bandwidth constraints~\cite{gladden_lessons_2024}.

To address the capacity limitations of interplanetary networks (IPN), free-space optical (FSO) communication is preferred over traditional RF communication. 
Optical links provide significantly higher data rates, ranging from Gbps to Tbps, while also requiring lower transmission power due to their narrow beam spread~\cite{wang_free_2024}.
As a result, optical terminals can be integrated into a satellite platform, reducing size, weight, power, and cost (SWaP-C) compared to RF systems.
The SWaP-C criteria are critical for space missions, thus making laser communications particularly advantageous.

\begin{figure}[t]
    \centering
    \includegraphics[width=1.0\linewidth,height=\textheight,keepaspectratio]{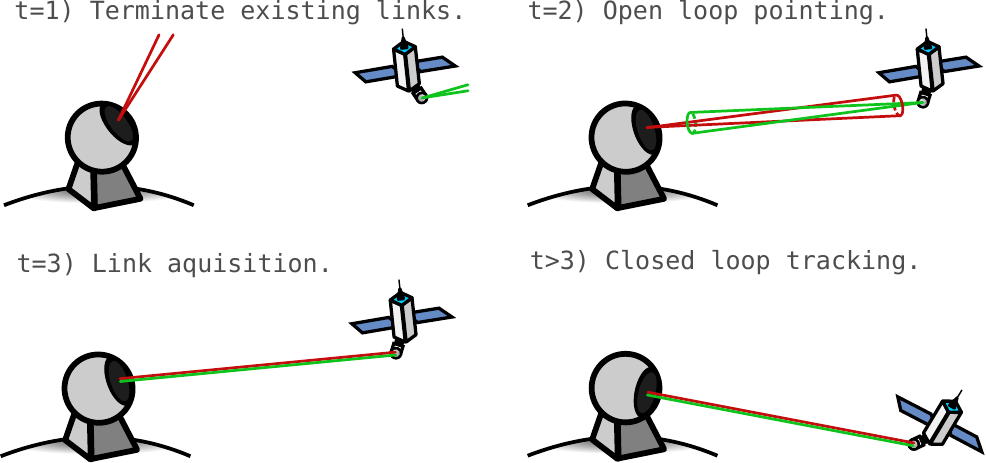}
    \caption{Pointing, acquisition, and tracking (PAT) between two free-space laser terminals. After the optical terminals terminate their existing link at $t=1$, open-loop pointing for the new contact window begins at $t=2$. The optical heads slew toward each other using coarse pointing assemblies (CPA) or through spacecraft body pointing. At $t=3$, link acquisition begins when one of the optical terminals illuminates the other terminals' field-of-view (FOV) sensors. For $t>3$ and until the end of the contact window, the terminals continue tracking each other over their orbits to maintain the optical link.}
    \label{fig:pat_process}
\end{figure}

Optical links establish and maintain connectivity through the pointing, acquisition, and tracking (PAT) process~\cite{bhattacharjee_ondemand_2024}. 
Establishing an optical link requires precise mechanical alignment between the transmitting laser and the receiving telescope~\cite{wang_free_2024}. 
Figure~\ref{fig:pat_process} illustrates the multi-step sequence required to form an optical link between a ground station and a satellite. 
A major challenge in this process is the delay introduced by pointing and acquisition, which must be completed before any data can be transmitted. 
This process, known as retargeting, refers to the mechanical act of pointing an optical terminal’s laser from its current partner to the next receiver. 
Retargeting delay includes coarse pointing and link acquisition, and can last from several seconds to minutes, depending on hardware and link geometry.

The precise pointing requirements and retargeting delays of optical terminals require link opportunities to be scheduled in advance. 
Interplanetary optical links must also retarget frequently because backhaul resources are limited, spacecraft have tight energy budgets, and orbital dynamics continually disrupt line-of-sight. 
Contact plan design (CPD) addresses this problem by algorithmically selecting which links connect terminals, thereby maximizing specific properties of the resulting contact plan. 

Recent work has proposed autonomously scheduling interplanetary optical links between Earth and Mars to overcome the limitations of manual scheduling~\cite{gerard_autonomous_2024}. 
The authors used a greedy backhaul scheduling algorithm to compute a contact plan that maximizes backhaul network capacity. 
The scheduling results demonstrated the effectiveness of a maximum-flow capacity model, as shown through simulations of multiple scenarios with varying constellation sizes.

A PAT-aware CPD framework can more accurately represent the usable portion of each contact window, reducing the delta between the scheduled capacity and the capacity actually delivered once the plan is executed. 
We also hypothesize that a linear programming (LP) model can further increase backhaul capacity by evaluating link sequences over the entire planning horizon rather than making local decisions at each time step. 
Greedy and heuristic approaches lack this global view and therefore cannot account for how current scheduling choices restrict or enable future topologies. 
This limitation is especially problematic in temporal networks, where link opportunities depend on orbital motion and retargeting behavior. 
An LP scheduler can capture these dependencies, but at the cost of higher computational complexity and reduced scalability. 

\subsection*{Contributions}

This work makes several contributions that advance the state of optical interplanetary networking beyond heuristic scheduling and simplified link models. 
First, we develop a temporal network capacity model that captures directional flows, multi-hop relay behavior, and dynamic topology within a time-expanded representation that preserves all feasible source–destination journeys. 
By explicitly computing the effective contact duration using the retargeting delay, the model provides an accurate foundation for analyzing optical backhaul performance, enabling future research to reason about link availability, resource contention, and contact sequencing with fidelity not achievable using static or zero-delay abstractions. 
Building on this model, we introduce a MILP scheduler that yields not only higher capacity but also a model that accounts for realistic optical terminal constraints, such as multi-terminal spacecraft and asymmetric bit rates. 
To quantify system-level behavior, we introduce an optical network duty-cycle metric that reveals how much of a scheduled window is spent transmitting versus retargeting, thereby exposing performance bottlenecks that previous models could not observe. 
Using realistic orbital dynamics, link budgets, and modem configurations informed by NASA’s DSOC, LCRD, and TBIRD programs, we verify that the PAT-aware MILP achieves over 30\% more scheduled throughput than greedy approaches and fundamentally shifts optimal behavior toward fewer but longer contacts. 
Together, this work contributes to establishing a rigorous, retargeting-aware foundation for autonomous optical link scheduling.
This work can be leveraged for multi-mission, self-optimizing deep-space networks capable of meeting the data demands of next-generation planetary exploration. 

\section{Free-Space Optical Communication}

\begin{figure*}[!t]
    \centering
    \includegraphics[width=1.0\linewidth,height=\textheight,keepaspectratio]{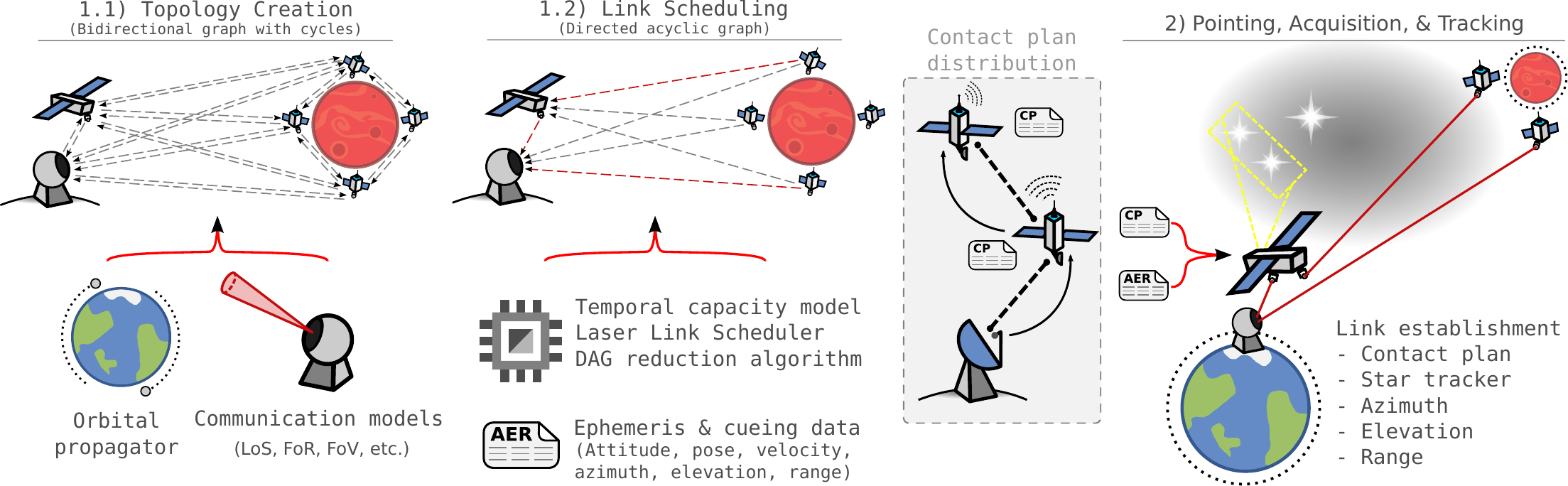}
    \caption{Framework for OISL scheduling and PAT in deep space DTNs. \textbf{1.1)} The contact plan (CP) is computed using orbital propagators and communication models. \textbf{1.2)} Topology transformations are applied, such as the DAG topology reduction, and the optical link contact plan is scheduled using a scheduling algorithm, such as the Laser Link Scheduler (LLS).
    Mission-critical, delay-intolerant data can then be manually added to the contact plan before it is distributed to the DTN nodes. \textbf{2)} Pointing, acquisition, and tracking (PAT) for free-space laser link establishment uses the scheduled contact plan, azimuth, elevation, range report, and the satellite's onboard star trackers. After establishing the optical links, the route tables can be computed using a DTN routing algorithm. Nodes query the route table for the outbound queue used to forward the bundle.}
    \label{fig:contact_plan_system_model}
\end{figure*}

In the following section, we describe how PAT and link scheduling through CPD are handled to establish and maintain connectivity for optical links over interplanetary distances.

\textit{Pointing, Acquisition, and Tracking (PAT)}. When open-loop pointing begins between two optical terminals, the optical heads slew toward each other using the two-axis gimbal in their coarse pointing assemblies (CPA) or spacecraft body pointing.
Ephemeris positioning and cueing data are computed through star trackers, GNSS/GPS, and two-line element (TLE) set orbital propagators.
To initiate link acquisition, one of the optical terminals must illuminate the other satellite's field-of-view (FOV) sensor using the onboard high-divergence-angle beacon beam to guide alignment of the optical terminal.
This common technique makes beam-searching algorithms more efficient, as the beacon's wide spot beam is much easier to find and track than the narrow laser beam~\cite{fernandez-nino_rfassisted_2024}.
Another technique, developed by NASA's Jet Propulsion Laboratory (JPL), is a star tracker-based pointing method to remove the need for a beacon~\cite{swank_beaconless_2016}. 
Deep-space link acquisition must be done without any direct feedback from the partner terminal, as the one-way light time can often exceed the contact window duration.
Deep-space contacts, for example, between Earth and Mars, may be on the order of 30 minutes, with a one-way light time of 40 minutes, depending on the orbit~\cite{suhardiman_simulation_2025}.
At the end of the link acquisition, a feedback loop is established through closed-loop fine-pointing using the fast steering mirrors (FSM).
For the remainder of the contact window, the optical terminals continue closed-loop tracking of each other over their orbital trajectories, utilizing point-ahead to maintain the optical link.

\textit{Link Scheduling.} A contact plan is a time-varying network model, formatted as a list of fields describing each contact opportunity.
Satellites and ground stations use the contact plan to synchronize when to establish optical links and with which partner terminals.
In near-Earth deployments, synchronization can be achieved in real time through a software-defined network (SDN)~\cite{barritt_loon_2018}.
However, in deep-space networks, the synchronization between satellites for PAT must be achieved well before the link is established, accounting for the propagation delay between the planets and the link setup duration.
CPD can optimize link selection, thereby maximizing backhaul network capacity by scheduling optimal flows across the network topology~\cite{akrida_temporal_2019, dhara_mfr_2022, gerard_autonomous_2024}.
Furthermore, fair contact plans will maximize resource utilization in a network while balancing source-to-destination node transmission, regardless of the routing algorithms used~\cite{fraire_design_2014}.
Currently, both manual~\cite{gladden_lessons_2024} and automated~\cite{damiani_automated_2006} CPD methods are used, such as ESA's ESTRACK scheduler.
However, the long propagation delay between planets imposes a hard upper limit on how dynamically a system can respond to network feedback. 
A scheduled contact plan should only contain transmissions in the same frequency band; for example, optical and RF transmissions are orthogonal and will not interfere with each other, so they should be scheduled separately.
Mission-critical telemetry, tracking, and command (TT\&C) data is typically exchanged through a low-gain RF antenna on the spacecraft bus. 
This avoids atmospheric attenuation and eliminates the need for precise antenna or body pointing during routine operations. 
In contrast, scientific telemetry, the dominant contributor to deep-space data volume, flows through the optical terminal payload and is therefore scheduled within the laser contact plan. 

\section{Related Work}

In the following, we present various CPD approaches developed for FSO and RF-based communications in LEO constellations and in interplanetary scenarios. 
These works provide the foundation for defining the optical contact plan design problem and formulating both heuristic algorithms and LP solutions.

\subsection{RF Link Scheduling} 
For RF-based LEO constellation scheduling, one of the earliest works introduced the Fair Contact Plan (FCP), a scalable polynomial-time algorithm designed to maximize fairness in DTN~\cite{fraire_design_2014}.
Initially formulated as a MILP problem, the authors demonstrated that contact scheduling quickly becomes intractable for larger networks, leading to the development of a heuristic algorithm over a time-expanded graph to compute feasible schedules.

The authors in~\cite{fraire_trafficaware_2016} introduced constraints in the model formulation for storage and transmission capacity, which reduces system resource usage by 42\% compared to existing work.
However, the increasing complexity of LP model constraints highlights the need for heuristic-based scheduling approaches to scale effectively in larger constellations.
Similarly, the authors in~\cite{fraire_scalability_2021} explored the scalability limitations of LP scheduling models that incorporate realistic constraints, such as battery dynamics, flow control, and interface restrictions.  
These constraints significantly impact tractability, particularly for large satellite constellations, where an upper bound in computational feasibility is reached at around 50 nodes.
A MILP problem can be reformulated into a pure LP by relaxing or removing integer constraints, allowing the LP model to solve much larger problem spaces.
However, this simplification alone may compromise the feasibility of the solution.
For instance, interface constraints can be relaxed in RF communication due to broadcasting and collision resolution at the MAC layer. 
However, this is not feasible for optical systems requiring precise pointing and alignment.
The trade-off between computational complexity and scheduling performance underscores the need for scalable yet practical solutions that account for the unique constraints of interplanetary networking.

FSO poses a unique challenge for routing and scheduling optimization in interplanetary satellite networks.
For instance, while the work in~\cite{dhara_mfr_2022} demonstrates topology-pruning techniques to manage the computational complexity of maximum flow calculations, and the work in~\cite{elalaoui_mars_2020} introduces multi-attribute decision-making for bundle scheduling, these solutions primarily focus on data flow optimization without considering the precise PAT required for optical links.
When considering these methods with an FSO physical layer, it is unclear how PAT delay affects the resulting schedules and route table computation.

\subsection{Optical Link Scheduling}
Interplanetary contact scheduling requires balancing computational efficiency and algorithmic performance, as highlighted in various scheduling approaches.  
The greedy backhaul scheduling algorithm, introduced in~\cite{gerard_autonomous_2024}, utilizes the blossom graph theory algorithm to compute a maximal matching at each time slice, thereby maximizing the increase in network capacity.
However, because it is heuristic, the algorithm does not guarantee a globally optimal solution across the entire temporal graph, potentially leading to underutilized network resources despite its computational efficiency.  

Similarly, existing research on dynamic inter-satellite links examines energy efficiency and latency optimization but often overlooks the fundamental distinction between optical link scheduling and real-time routing decisions.
For example, the work in~\cite{bhattacharjee_ondemand_2024} does not account for the fact that the route table can only be computed once the optical link topology is established.
In this way, the routes forwarding traffic through a given satellite must be constrained by the number of optical terminals.
The separation between contact scheduling and routing is crucial because free-space laser communications require precise synchronization of node contact opportunities. 
Routing algorithms alone cannot solve the link scheduling problem, because the optical link topology must be computed during the planning phase before real-time routing occurs.

There has been an increasing focus on improving LEO optical inter-satellite link (OISL) mesh topologies by minimizing hop count and data delivery delay~\cite{wang_federated_2024, ron_timedependent_2025, wang_intersatellite_2022}.
These solutions employ new algorithms, such as a degree-constrained minimum spanning tree, tailored for optical links, ensuring that each optical terminal only has a single partner terminal~\cite{nardin_contact_2021}.
While these methods provide valuable insights into contact scheduling via heuristic and MILP-based approaches, they are largely tailored to single-constellation networks with short propagation delays and minimal disruptions.
However, optical interplanetary networking presents unique challenges, including significant propagation delays, highly dynamic topologies, and stringent pointing requirements.
These fundamental differences prevent the direct application of LEO constellation scheduling techniques to optical IPN link scheduling.
Nonetheless, the methodologies developed for LEO constellations are a strong foundation for modeling CPD in optical IPN.

Unlike static network design problems, OISL networks are constrained by link locality, satellite mobility, and dynamic topology changes, making conventional approaches unsuitable~\cite{bhattacherjee_network_2019}.
Given the limited number of optical terminals per satellite, traditional methods such as LP, random regular graphs, and ant colony optimization become intractable as the constellation scales.  
The authors in~\cite{bhattacherjee_network_2019} introduced a graph theory approach to optimize topology formation using motifs.
Motifs reduce computational complexity by grouping multiple OISLs into a single computation unit, preventing exponential growth in problem size.
While motifs have not been evaluated with interplanetary OISLs, the concept lends itself well to the directed topology of links between source and destination planets. 

The authors in~\cite{tian_distributed_2022} introduced a satellite link allocation algorithm based on a non-dominated sorting genetic algorithm, demonstrating reduced communication delays across five simulated cislunar communication scenarios. 
However, the algorithm does not account for optical link constraints, such as the presence of a single partner optical terminal, nor does it consider the impact of link setup and retargeting delays on the total end-to-end communication delay.

A link scheduling and routing algorithm using an online Lyapunov optimization with local buffer occupancy is proposed in~\cite{ren_research_2024}.
While this algorithm improves the bundle delivery ratio, it is impractical in IPN because it relies on real-time buffer information.
By the time the route tables arrive at the deep-space satellites and the optical links are retargeted, the buffer occupancy and, thus, the optical link schedule will differ.

PAT-aware CPD has not yet been explored in the existing literature.
To the best of our knowledge, there is not a single CPD framework that accounts for interplanetary disruptions, long propagation delays, optical interface limits, and the retargeting delays introduced by PAT. 
This leaves a gap between scheduling models in the literature and the operational behavior of current and future optical backhaul networks. 
The growing interest in high-rate optical communications makes this gap increasingly important to address. 
A CPD framework that incorporates PAT constraints is therefore needed to support accurate performance analysis and the design of autonomous optical networks for future space missions. 

\section{Temporal Capacity Model} \label{sec:backhaul_model}

The following section introduces the graph data structure used to model the temporal network, along with a capacity model that leverages the network's directionality.
It does this by defining flow-based capacity from the source nodes that generate data to the sink nodes located on Earth. 
This capacity estimate serves as an edge weight for routing or, as in this work, for scheduling contacts in a CPD framework. 

\subsection{Time-Extended Graph}

Terrestrial networks are typically modeled as static directed graphs, where vertices represent communicating entities and edges correspond to communication opportunities between them.  
Formally, a static network is a directed graph $G = (V, E)$, where $V$ is a set of $n$ vertices (nodes) and $E \subseteq V \times V$ is a set of $m$ directed edges (links).  
Each edge $e = (u,v) \in E$ has an associated capacity $c_e \geq 0$, representing the maximum flow that can be carried from $u$ to $v$.  
We further designate a set of sources $S \subseteq V$, which are flow entry points, sinks $T \subseteq V$, which are flow endpoints, and relays $R \subseteq V$, which are vertices in between source and sink nodes.
There may exist a path between a source and sink node that does not cross any relay node.

The network topology under study is a deep-space optical backhaul network modeled after existing missions.
In contrast to most terrestrial networks, space communication networks are inherently temporal: the availability of a link depends on the orbital positions of the spacecraft, leading to connectivity that changes over time.  
To capture this dynamic, we subdivide the time horizon into $K$ discrete and equally sized time steps, through a process called graph fractionation, and construct a time-extended graph (TEG), denoted $\mathit{TEG}=(V', E')$.  
Here, $V' = \{ v_k : v \in V,\, k=0,\dots,K \}$ represents time-stamped copies of each vertex, and for every $(u,v)\in E$ available at time $k$, we introduce a time-edge $(u_k,v_k)\in E'$ with capacity $c_{(u_k,v_k)}$.  
This construction yields a static representation of the temporal network in which flows respect both edge capacities and causality across discrete time steps.  

\begin{figure}[t]
    \centering
    \includegraphics[width=1.0\linewidth,height=\textheight,keepaspectratio]{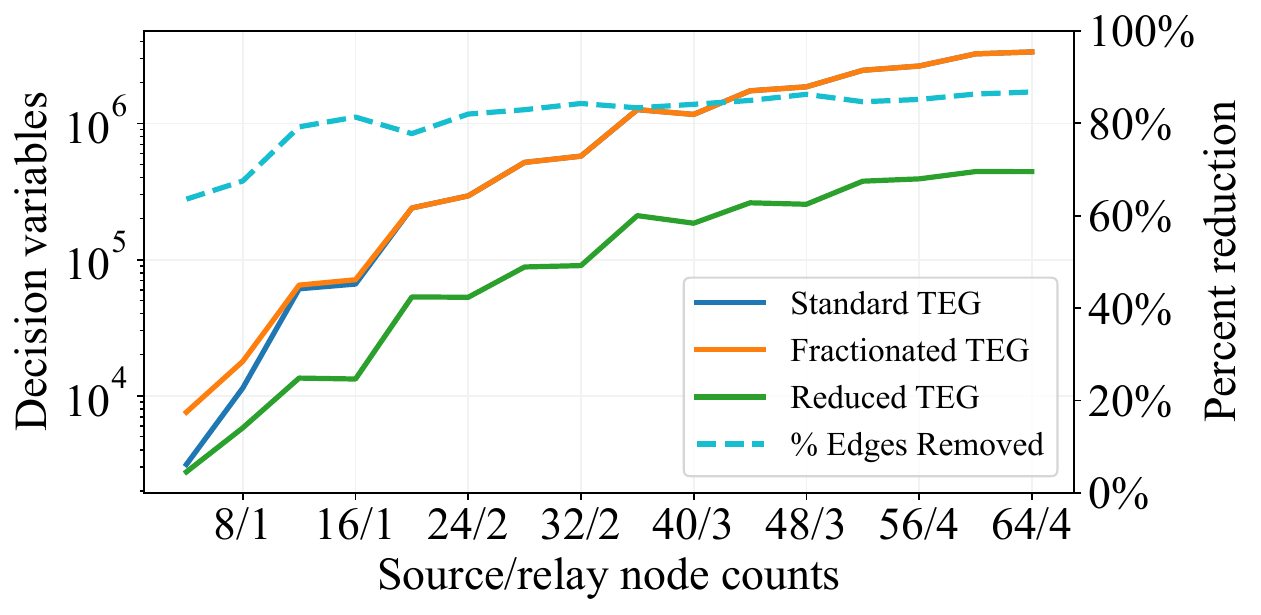}
    \caption{Number of decision variables after reduction.}
    \label{fig:decision_variable_count}
\end{figure}

One challenge with the TEG data structure is its size: because a copy of the graph is created for each time step, the worst-case number of edges can grow to $O(K|V|^2)$.
The blue line labeled Standard TEG in Fig.~\ref{fig:decision_variable_count} shows the exponential growth of decision variables as the number of nodes in the TEG increases.
This rapid growth significantly impacts the scalability of scheduling, routing, and optimization problems, which typically scale with the number of edges, rendering them quickly intractable~\cite{fraire_scalability_2021}.

We therefore apply a pruning step that reduces TEG while preserving all feasible temporal $s$--$t$ paths (also known as journeys).  
The pruning eliminates edges that cannot contribute to capacity; these include edges that do not flow towards a sink node, such as redundant bidirectional links, and edges that connect a source to another source, $e = (u,v) \in E$ such that $u,v \in S$.
By reformulating the topology as a directed acyclic graph (DAG), as shown in Fig.~\ref{fig:contact_plan_system_model} (step 1.2), the resulting pruned TEG contains only edges that participate in valid one-hop, $s$--$t$, or two-hop, $s$--$r$--$t$, paths.  

Figure~\ref{fig:decision_variable_count} shows the number of decision variables, which scales with the number of edges in the TEG. 
Fractionating the TEG normalizes the contact window durations across time slices, which increases the number of variables at lower node counts and has been shown to improve fairness in CPD algorithms~\cite{fraire_design_2014}. 
Applying the DAG reduction then removes edges that cannot participate in any valid temporal journey, reducing the number of decision variables by an order of magnitude while preserving all feasible max-flow paths. 
The dashed blue line highlights that approximately 80\% of edges can be pruned when transforming the standard TEG into the reduced TEG. 

\subsection{Temporal Max-Flow Min-Cut Theorem}

The maximum–flow minimum-cut theorem states that for a finite directed network $G=(V, E)$ with nonnegative capacities and distinguished vertices $s,t\in V$, the maximum value of any $s$--$t$ flow equals the minimum capacity of any $s$--$t$ cut.
Our model builds on the max–flow min-cut theorem, extended from static graphs to temporal flows by representing the temporal network as a TEG of polynomial size~\cite{akrida_temporal_2019}.  
Given a temporal digraph $G=(V, E, K)$ with capacities $c_e \ge 0$ and unbounded node buffers ($B_v=\infty$), the maximum temporal flow equals the capacity of a minimum temporal $s$--$t$ cut, and every feasible temporal flow decomposes into flows along $s$--$t$ journeys (i.e., paths with strictly increasing labels).  
Note that an algorithm that is optimal on each static snapshot of the network may be globally suboptimal across the temporal graph.
This is because choices that maximize instantaneous $s$--$t$ flow can consume capacities at labels that preclude any valid $s$--$t$ journey later.

\subsection{Edge Flow}
The theorem, in the case of unbounded node buffers, serves as the basis for our interplanetary backhaul capacity model, where the flow capacity of any given edge between nodes $u$ and $v$, across all $K$ states, is defined by
\begin{equation} \label{eq:flow}
    \mathit{flow}_{e_k} = \sum_{k=1}^{K} L_{e_k} \cdot T^{\mathit{eff}} \cdot B_{e},
\end{equation}
where $L_{e_k} = L_{(u,v)_k}$ is a decision variable set to 1 if the edge was selected in time step $k$ and 0 if it was not. $T^{\mathit{eff}}$ is the effective contact window duration, defined as $T_{k} - T_{\mathit{retarget}}$, the difference between the duration of the contact window and the retargeting delay, which is further defined in~\cite{gerard_modeling_2025}.
The data rate of the flow, $B_{e}$, measured in bits per second, is defined as $\min(B_u, B_v)$, the minimum data rate between the transmitting and receiving modems. 

\subsection{Capacity Model}

\begin{figure}[t]
    \centering
    \includegraphics[width=0.8\linewidth,height=\textheight,keepaspectratio]{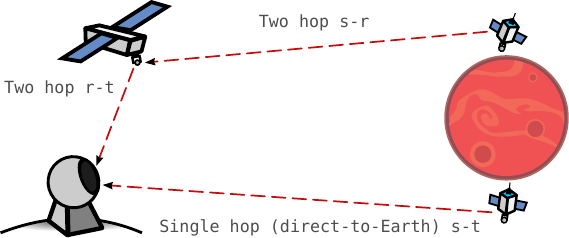}
    \caption{System model of both types of temporal paths considered, single-hop direct-to-Earth and two-hop source, relay, destination paths.}
    \label{fig:optical_link_path_types}
\end{figure}

By recognizing that this is a specialized network rather than a general-purpose one, we can reduce the topology to a subset of paths and compute the total capacity based on the inflows and outflows at each node.
Data can flow along two different types of temporal paths, shown as a snapshot in Fig.~\ref{fig:optical_link_path_types}.
The first type is a single-hop direct-to-Earth (DTE) path, in which a deep-space satellite has a direct link to an optical ground station on Earth.
The second type is a two-hop path, where the deep-space satellite has a link to a relay satellite orbiting Earth, and the relay satellite has a second link to an optical ground station at a much higher data rate.
DTE paths are simpler, requiring only a single link, but they experience more disruptions due to orbital occultations and atmospheric attenuation.
On the other hand, the two-hop indirect path is limited by the size of the relay satellite's receive telescope and the power of the deep-space modem and optical amplifier.
Two-hop paths use DTN Earth relay nodes to lift the topology from a temporal bipartite graph into a directed acyclic graph (DAG).
This has the practical benefit of leveraging the relay nodes to effectively extend the contact window of the ground stations, which would otherwise have no contact opportunities when facing away from a source planet.

An edge that increases the flow capacity must be in one of the following categories:
\begin{enumerate}
    \item The edge corresponds to a one-hop path from source to destination.
    \item The edge is from a relay node to a destination node and is part of a two-hop path, from source to relay to destination.
    \item The edge is from a source node to a relay node, the relay node is orbiting the same planet as the source node or the destination planet, and the edge is part of a two-hop path from source to relay to destination.
\end{enumerate}

Eliminating network paths reduces the problem's complexity, enabling the LP model to scale to larger networks.
This restriction on possible paths is validated based on existing missions that use single-hop direct-to-Earth links or two-hop paths with a single relay node, either orbiting the source planet or the destination planet.
The minimum value between the two for each node is the minimum cut or bottleneck.
The flow capacity of a node $v$, $C_{v}$, is the balance between the inflow of data from other nodes and the outflow of data to other nodes.
Therefore, $C_{v}$ is defined as
\begin{equation} \label{eq:capacity}
    C_{v} = \min(\mathit{inflow}_v, \mathit{outflow}_v),
\end{equation}
where $\mathit{inflow}_v$ for a relay node optical ground station is defined as $\sum \mathit{flow}_{u,v} \forall u \in V'_{\mathit{source}}$, the summation of all flows from a source node, $u$, directed to relay node $v$.
The $\mathit{outflow}_v$ for an Earth relay node is calculated as $\sum flow_{v,u} \forall u \in V'_{\mathit{sink}}$, i.e., the summation of all flows forwarded from the relay node $u$ to any destination ground station $v$.
The $\mathit{outflow}_v$ for a ground station is set to $\infty$ since it is a destination node.
This capacity model can then be used to schedule contact plans or compute route tables that maximize the total satisfiable demand across the network.

\section{Optical Contact Plan Design} \label{sec:mip_model}

To maximize network capacity, the objective function should maximize the sum of the flow-based capacities of each relay satellite and the destination optical ground station. 
The MILP model is based on Eq.~\ref{eq:flow}, which represents temporal capacity along an edge, and is formally defined as follows,

\begin{alignat}{3}
  & \text{max } & & \sum_{v}^{V'} C_{v} \label{eq:objective_fn} \\
  & \text{s.t. } & \hspace{0.5em} & C_{e_k} \le L_{e_k} \cdot T_{k} \cdot B_e & \hspace{0.3em} &\forall e_k \in E' \label{eq:big_m_edge_cap} \\ 
  &                     & \hspace{0.5em} & C_{e_k} \le T^{\mathit{eff}} \cdot B_e & \hspace{0.7em} &\forall e_k \in E' \label{eq:big_m_edge_cap_eff} \\ 
  &                     & \hspace{0.5em} & \mathit{inflow}_{v} \le \sum_{u}^{V'_\mathit{source}} \sum_{k}^{K} C_{(u,v)_k} & \hspace{0.3em} &\forall v \in V'_{\mathit{relay}} \label{eq:inflow_relay} \\
  &                     & \hspace{0.5em} & \mathit{outflow}_{u} \le \sum_{v}^{V'_\mathit{sink}} \sum_{k}^{K} C_{(u,v)_k} & \hspace{0.3em} &\forall u \in V'_{\mathit{relay}} \label{eq:outflow_relay} \\
  &                     & \hspace{0.5em} & \mathit{inflow}_{v} \le \sum_{u}^{V'} \sum_{k}^{K} C_{(u,v)_k} & \hspace{0.3em} &\forall v \in V'_{\mathit{sink}} \label{eq:inflow_gs} \\ 
  &                     & \hspace{0.5em} & \sum_{v}^{V'} L_{(u, v)_k} \le 1 & \hspace{1.3em} &\forall u_k \in V' \label{eq:ic_optical} \\ 
  &                     & \hspace{0.5em} & \mathit{ECT}_{u} \ge \mathit{ECT}_\mathit{AVG} \cdot \epsilon & \hspace{0.3em} &\forall u \in V'_\mathit{source} \label{eq:fairness} 
\end{alignat}

\textit{Objective Function.} The MILP model aims to maximize the objective function, defined in Eq.~\ref{eq:objective_fn}, as the summation of capacity variables defined for the Earth relay nodes, Eq.~\ref{eq:inflow_relay} and Eq.~\ref{eq:outflow_relay}, and the optical ground station nodes, Eq.~\ref{eq:inflow_gs}.

\textit{Capacity Constraints.} The core formulation of the model is based around the capacity constraints starting from Eq.~\ref{eq:flow}, which defines the flow capacity for a specific edge. Equation~\ref{eq:capacity} then combines specific flows to give a capacity estimation.
The model then takes these to formulate a decision variable in Eq.~\ref{eq:big_m_edge_cap} and an upper bound on the actual capacity in Eq.~\ref{eq:big_m_edge_cap_eff}, which is what is actually maximized in the objective function, Eq.~\ref{eq:objective_fn}.
Equations~\ref{eq:big_m_edge_cap} and ~\ref{eq:big_m_edge_cap_eff} are a Big-M constraint reformulation of the effective capacity over each edge in the network, where these constraints bind the capacity variables.
The first constraint, Eq.~\ref{eq:big_m_edge_cap}, represents the upper bound of capacity on the edge if it is not selected based on $L_{e_k}$.
This constraint applies upward pressure on edge selection to stop the model from minimizing the retargeting delay, defined by Eq.~\ref{eq:model_retarget}, by selecting no edges.
The second constraint, Eq.~\ref{eq:big_m_edge_cap_eff}, imposes the retargeting delay on the total effective contact time of the edge.
These set the upper bound on an edge's capacity if the edge is selected.

The first two constraints impose an upper bound on network flow, including the reduction in effective contact time due to the retargeting delay, while maintaining the program's linearity.
Without this, the model would become non-linear, requiring far more processing to converge on a solution.
The relay satellite inflow (Eq.~\ref{eq:inflow_relay}), the relay satellite outflow (Eq.~\ref{eq:outflow_relay}), and the optical ground station inflow (Eq.~\ref{eq:inflow_gs}) aggregate the capacity over their corresponding edges and bind them to the backhaul model rules defined in section~\ref{sec:backhaul_model}.
These variables are ultimately the output variables used in the objective function to maximize network capacity.

\textit{Laser Interface Constraints.} Equation~\ref{eq:ic_optical} represents the interface constraints on the maximum matching for the lasers, where $\mathit{EDGES}=1$.
The model improves the interface constraints in existing work to support multiple optical terminals per node.
Each node may have multiple optical terminals, which all contribute to the same node-level capacity but have their individual edges.
By decoupling the decision variable for each node and its optical interfaces, we can also compute each terminal's retargeting delay independently.

\textit{Retargeting Delay Constraints.} The PAT delay model, introduced in~\cite{gerard_modeling_2025}, defined as
\begin{equation} \label{eq:model_retarget}
    T_\mathit{retarget} = T_\mathit{pointing} + T_\mathit{acq},
\end{equation}
is used in Eq.~\ref{eq:big_m_edge_cap_eff} to compute the retargeting delay of an optical link.
These values are then used as an upper bound on $T_{\mathit{eff}}$, binding the contact window to the dynamically computed effective contact duration.

\textit{Fairness Constraints.} The linear program implements fairness in Eq.~\ref{eq:fairness} through a soft constraint using enabled contact time (ECT), the duration that a node has been scheduled to transmit for in the contact plan.
The variable is loosely bound to the average ECT across all source nodes through $\epsilon$, which can be tuned based on the scenario.
For this work, we evaluated $\epsilon \in [0.5, 1.0]$ and found that a value of $\epsilon=0.95$ allowed the model to exchange some source node fairness to reduce retargeting delay and maximize capacity.
ECT for a given node is defined by
\begin{equation}
    \mathit{ECT}_{u} = \sum_{v}^{V'} \sum_{k}^{K} L_{(u,v)_k} \cdot T^\mathit{eff}.
\end{equation}

\section{Case Study: Optical Interplanetary Backhaul Network}\label{sec:evaluation}

In this section, we propose a duty cycle metric for optical networks, describe our evaluation methodology using realistic simulations and initial contact plans, and then evaluate multiple scheduling algorithms against each other in terms of network capacity and uptime.

\subsection{Optical Network Duty Cycle}

To quantify the effective utilization of an optical interplanetary network, we define the optical network duty cycle, $\mathit{ODC}$.
Let $L$ denote the set of temporal edges selected by the scheduling algorithm, where each edge $e=(u,v,k)$ corresponds to a pair of nodes $(u,v)$ during the $k$-th contact window of duration $T_k$.
Let $T^{\mathit{eff}}_{u,v,k}$ be the effective contact window duration.
The duty cycle of the network over all selected edges $L$ is defined as the ratio of contact time with the optical link up to the total contact window
\begin{equation} \label{eq:optical_dc}
\mathit{ODC} = 100 \cdot \frac{\sum_{(u,v,k) \in L} T^{\mathit{eff}}_{u,v,k}}{\sum_{k \in L} T_k}.
\end{equation}

This metric represents the proportion of contact time used for data transmission.
A contact plan with a duty cycle closer to 100\% indicates that the optical terminals spend less time retargeting and more time transmitting data.

When DTN relay nodes are introduced, the duty cycle can increase because multiple data sources can preload their data into the transmitting node’s buffer, thereby amortizing the cost of PAT, increasing the fraction of time spent transmitting.

\subsection{Simulation Setup}

\begin{figure}[t]
    \centering
    \includegraphics[width=1.0\linewidth,height=\textheight,keepaspectratio]{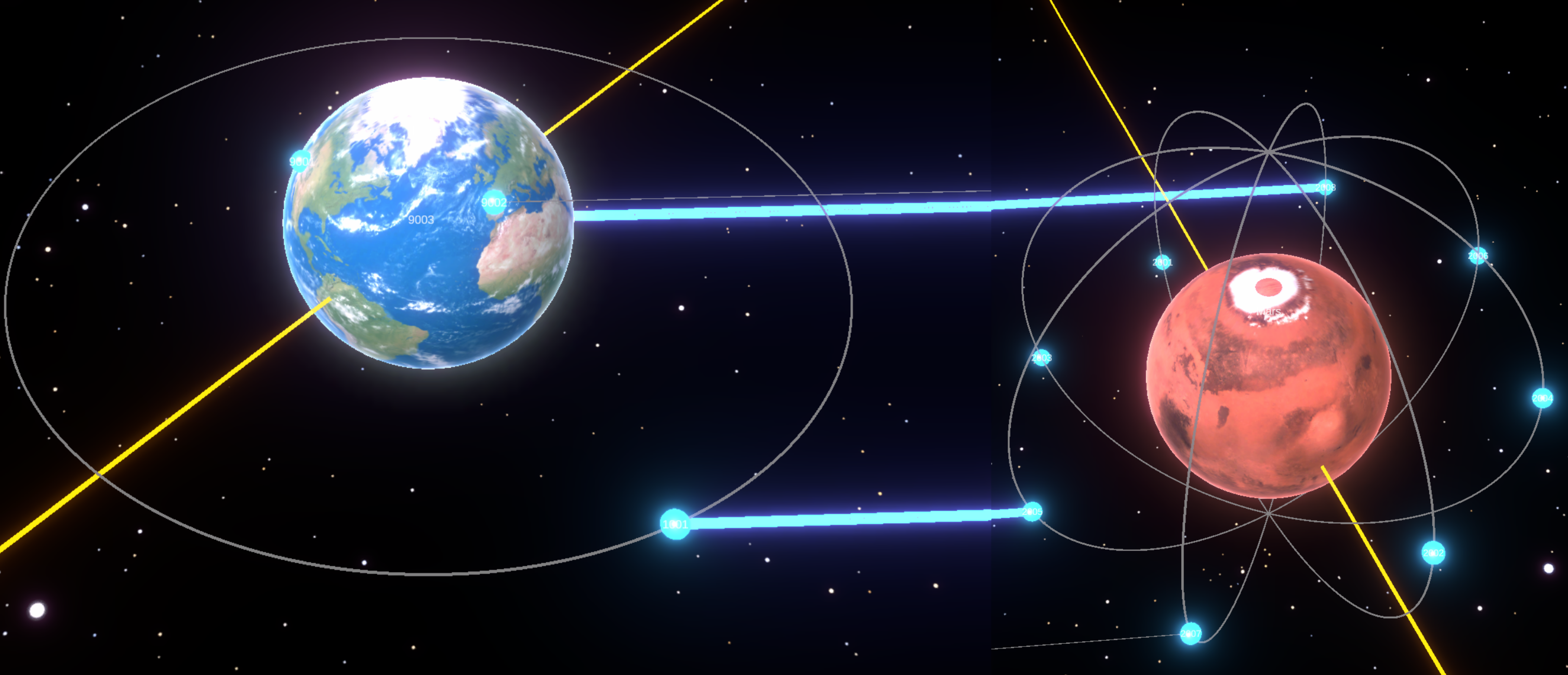}
    \caption{Mars-Earth interplanetary optical network (interplanetary link distance reduced for visualization) used in simulation to evaluate the CPD algorithms~\cite{bihan_ipnv_2024}.}
    \label{fig:simulation_network}
\end{figure}

The following section describes the simulation system model, shown in Fig.~\ref{fig:simulation_network}, its parameters, summarized in Tab.~\ref{tab:simulation_params}, and the evaluation methods used.

\begin{table}[t]
\centering
\caption{Simulation parameters~\cite{israel_early_2023, boroson_overview_2014, riesing_operations_2023, biswas_deep_2024}.}
\label{tab:simulation_params}
\begin{tabular}{@{}lll@{}}
\toprule
Parameter & IPN & LEO \\ \midrule
Modulation & PPM & DPSK \\
Bit rate & 50 mbps & 1.2 gbps \\
Link distance & 100 million km & 5,000 km \\
Slew rate Az & 1.0 deg/s & 2 deg/s \\
Slew rate El & 1.0 deg/s & 0.5 deg/s \\
FSM tip & 5.0 mrad/s & 8.5 mrad/s \\
FSM tilt & 5.0 mrad/s & 8.5 mrad/s \\
Dwell time & 0.5 s & 0.5 s \\
Beam width & 0.2 deg & 0.2 deg \\
FOU & 1.0 deg & 0.75 deg \\ \bottomrule
\end{tabular}
\end{table}

\textit{Optical Ground Stations.} The network contains three optical ground stations that use adaptive optics to account for atmospheric loss, as defined in NASA's LCOT standard~\cite{lafon_current_2023}.
Each ground station is positioned at one of the three NASA DSN sites~\cite{edwards_addressing_2024}, see the blue nodes on the surface of Earth in Fig.~\ref{fig:simulation_network}, and has a single optical terminal in the ground station housing.

\textit{Earth Relay Satellites.} The Earth relay satellites are each equipped with two optical terminals, as is the case for the NASA LCRD project~\cite{israel_early_2023}.
Figure~\ref{fig:simulation_network} shows a single one of these relay satellites orbiting Earth.
The relay satellites function as delay/disruption-tolerant (DTN) relay nodes that run the Bundle protocol. 
They aggregate, buffer, and take custody of data from deep-space satellites at a lower data rate until a contact opportunity with an optical ground station begins.
Upon contact with the ground station, it downlinks the buffered data. 
Using this approach, the TeraByte Infrared Delivery (TBIRD) mission can downlink 4.8 terabytes of data in a 5-minute contact window~\cite{riesing_operations_2023}.
The relay satellites are in a medium Earth orbit (MEO) at an altitude of 5,000 kilometers.
This orbit strikes a balance between higher data rates and ground-station visibility while maintaining a short orbital period, thereby allowing for more frequent contact opportunities. 
Additionally, the slower relative angular velocity makes it easier to establish optical links than with a LEO relay.
Earth-orbiting relays could be replaced with Mars-orbiting relays, as evaluated in~\cite{gerard_autonomous_2024}.

\textit{Deep Space Satellites.} The deep space satellites are modeled after those used in the NASA DSOC project~\cite{biswas_deep_2024}, specifically the Psyche mission~\cite{desoria-santacruz_systems_2024}.
The network comprises a satellite constellation in a four-plane Walker Star polar orbit around Mars, at an altitude of 500 kilometers, shown in Fig.~\ref{fig:simulation_network}.
Each spacecraft generates scientific data, which flows in a single direction towards Earth.
Transmissions between nodes must occur during a contact window only after a link is fully established through the PAT process shown in Fig.~\ref{fig:pat_process}.
Each deep space satellite orbiting Mars is equipped with a single optical terminal due to SWaP-C constraints, meaning they can only establish one link with one node at a time~\cite{desoria-santacruz_systems_2024}.

The most notable simulation parameters are of the modem's laser and digital signal processor (DSP).
The IPN links use pulse position modulation (PPM), a direct-detection method that offers high photon efficiency and enables energy-constrained operation in deep space~\cite{biswas_deep_2024}. These links have a data rate of 50 Mbps and are often over 100 million km.
The near-Earth links use differential phase shift keying (DPSK) modulation, which trades some of the photon efficiency of PPM for spectral efficiency, resulting in higher data rates at the cost of higher energy consumption~\cite{israel_early_2023}. These links have a data rate of 1.2 Gbps over a 5,000 km range.

To evaluate how well the scheduling algorithms scale, we consider scenarios with an increasing number of source and relay nodes and three optical ground stations.
The base scenario is shown in Fig.~\ref{fig:simulation_network}.
We compare the following scheduling algorithms across each scenario over 24 hours:
\begin{enumerate}
    \item \textit{LLS\_Greedy:} a graph theory-based model that leverages the max-flow network capacity model with the blossom algorithm to do an iterative maximal-matching, taking the greedy choice to maximize the capacity at each time step~\cite{gerard_autonomous_2024}. 
    This is the most recent state-of-the-art algorithm that maximizes capacity.
    This algorithm was updated from the previous work to support variable bit rates, multiple lasers per spacecraft, and pointing delay estimation.
    \item \textit{LLS\_Greedy (ZRK):} the same core algorithm as \textit{LLS\_Greedy} but with zero retargeting knowledge (ZRK), therefore it does not take retargeting delay into account for scheduling, contact time, or capacity.
    \item \textit{LLS\_MIP:} the proposed MILP scheduler using the max-flow capacity model.
    \item \textit{FCP:} a graph theory-based model that uses the duration each node is disabled for as a heuristic to perform a max weight maximal-matching optimizing the fairness of a constellation~\cite{fraire_design_2014}. This model was chosen as a baseline comparison from the literature that does not account for capacity.
    \item \textit{DTE Capacity:} a baseline scenario where only direct-to-Earth (DTE) links are used. This method provides the theoretical upper bound for ground station-only link capacity.
\end{enumerate}

Open-source implementations of the models in Python (NumPy, NetworkX, and Pulp libraries) using the Gurobi solver are publicly available\footnote{https://github.com/jason-gerard/laser-link-scheduler}.
The contact plans are generated using IPN-D\footnote{https://gitlab.inria.fr/jfraire/ipn-v}~\cite{bihan_ipnv_2024}.

\subsection{Discussion of Results}

\begin{figure*}[t]
    \centering
    \begin{subfigure}{0.32\textwidth}
        \includegraphics[width=\textwidth]{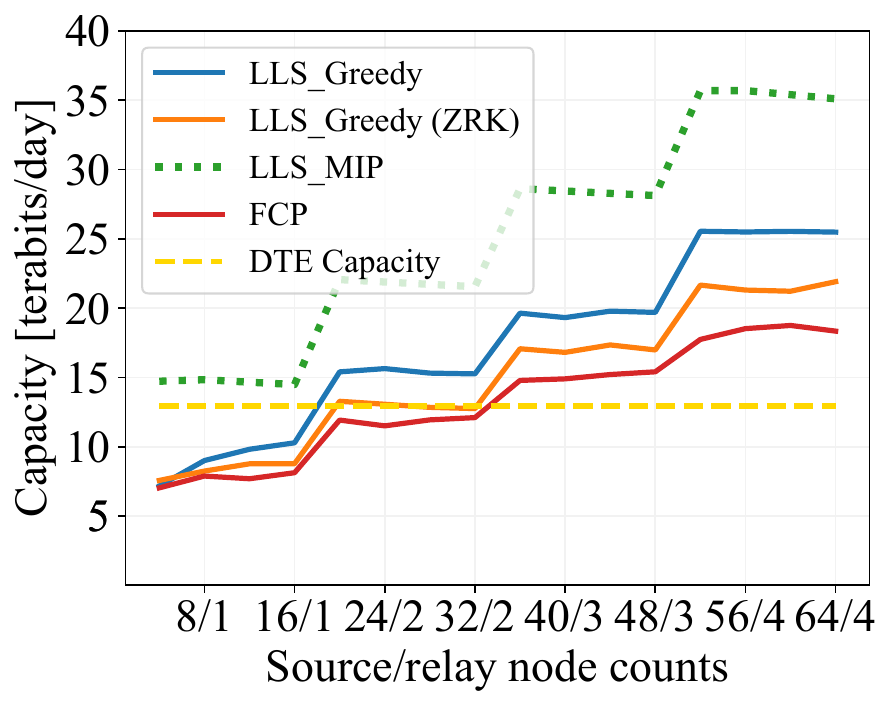}
        \caption{Network capacity}
        \label{fig:capacity}
    \end{subfigure}
    \begin{subfigure}{0.32\textwidth}
        \includegraphics[width=\textwidth]{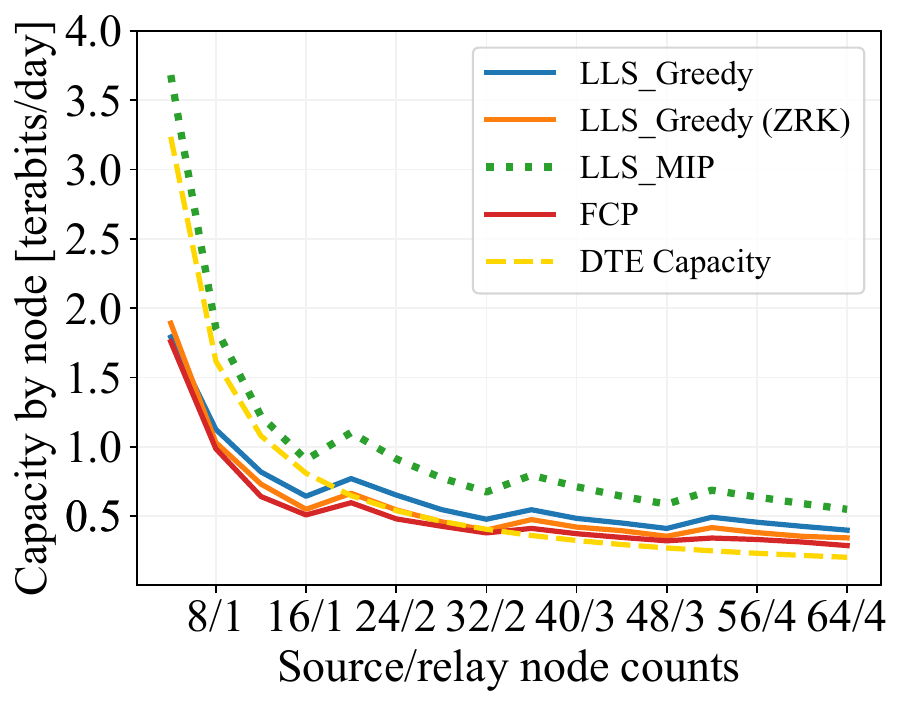}
        \caption{Node capacity}
        \label{fig:capacity_by_node}
    \end{subfigure}
    \begin{subfigure}{0.32\textwidth}
        \includegraphics[width=\textwidth]{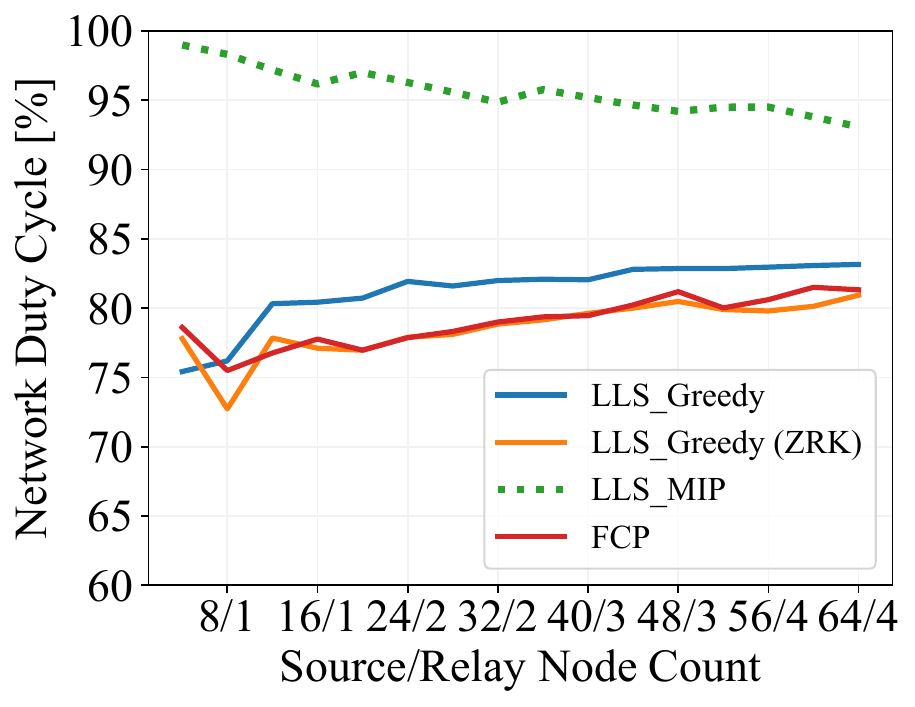}
        \caption{Optical network duty cycle}
        \label{fig:network_duty_cycle}
    \end{subfigure}

    \begin{subfigure}{0.32\textwidth}
        \includegraphics[width=\textwidth]{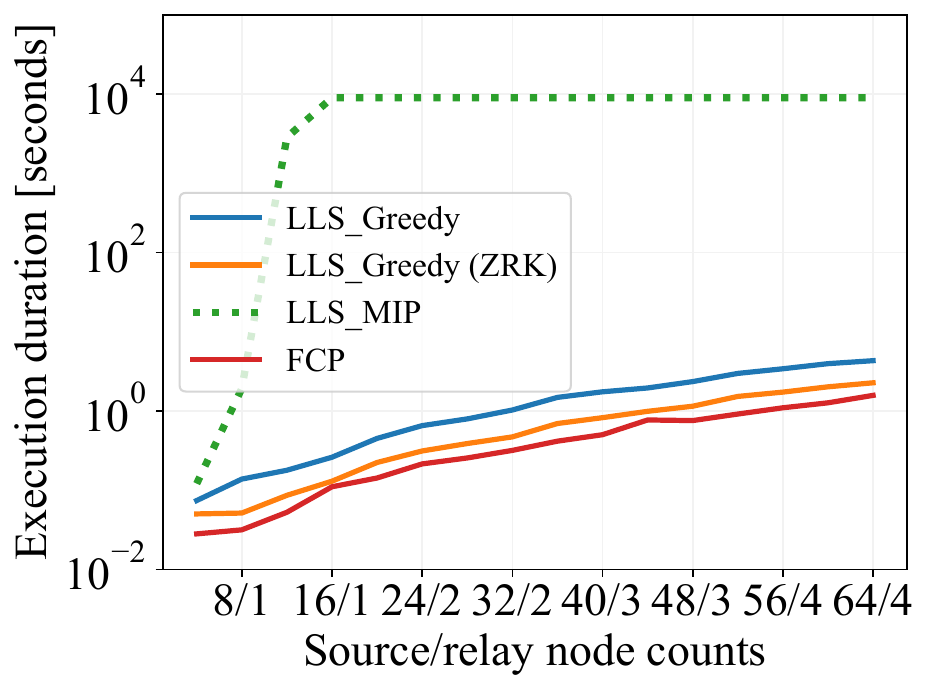}
        \caption{Algorithm runtimes, log scale}
        \label{fig:run_times}
    \end{subfigure}
    \begin{subfigure}{0.32\textwidth}
        \includegraphics[width=\textwidth]{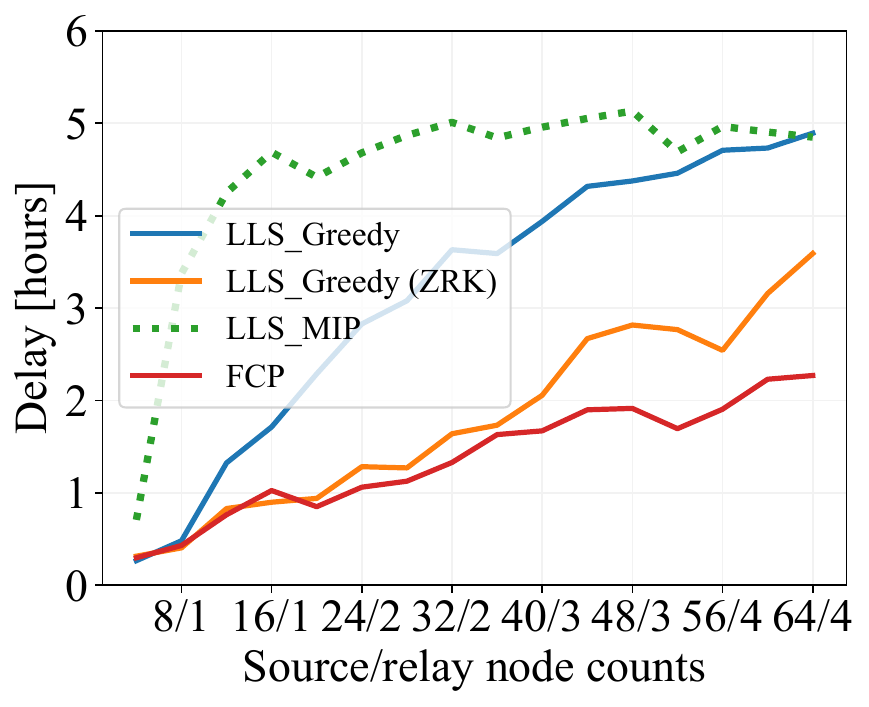}
        \caption{Average inter-contact delay}
        \label{fig:source_node_delay}
    \end{subfigure}
    \begin{subfigure}{0.32\textwidth}
        \includegraphics[width=\textwidth]{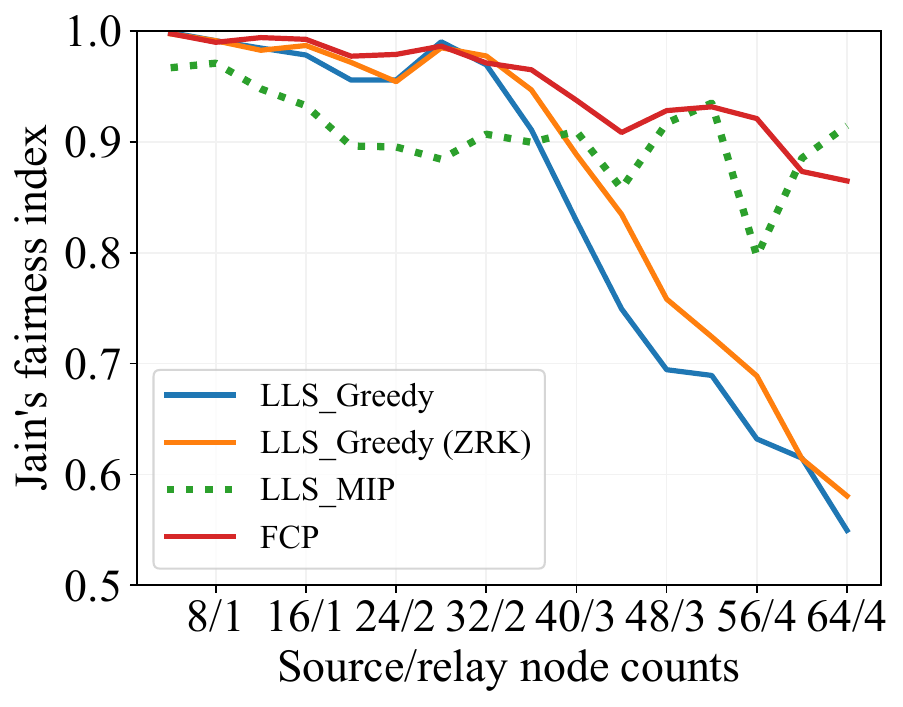}
        \caption{Scheduled fairness}
        \label{fig:scheduled_fairness}
    \end{subfigure}

    \caption{Performance comparison of algorithms across different-sized networks over 24 hours.}
    \label{fig:performance_comparison}
\end{figure*}

\textit{(a) Network Capacity.} Figure~\ref{fig:capacity} shows the scheduled network capacity, $C$, defined by $\sum_{v}^{V'_\mathit{sink}} C_{v}$, which measures the amount of data that can be backhauled to Earth within the scheduled contact plan $L$.
Scheduled capacity is the primary metric of concern when evaluating backhaul networks.
The MILP model, LLS\_MIP, has more than 30\% more capacity than the LLS\_Greedy algorithm across the different constellation sizes.
Although the branch-and-bound algorithm does not generally scale well as problem size increases, careful constraint selection allows the model to continue outperforming others even at larger network sizes.

\textit{(b) Node Capacity.} Figure~\ref{fig:capacity_by_node} illustrates the average data each node can transmit to Earth over 24 hours, providing insight into resource allocation as the constellation scales.
The relative trends among the algorithms remain consistent with those observed for overall network capacity.
The difference between the DTE and algorithm capacity highlights how effectively each algorithm leverages DTN relay nodes to increase capacity.
These types of nodes are fundamental to many IPN architectures~\cite {israel_lunanet_2020}.

\textit{(c) Optical Network Duty Cycle.} Figure~\ref{fig:network_duty_cycle} shows the operational duty cycle of the optical terminals across the interplanetary network based on Eq.~\ref{eq:optical_dc}.
The trends of this graph show the proportion of the scheduled contact time during which the optical terminal is transmitting data.
The higher the value, the less time the network spends retargeting the optical terminal performing PAT, and the more time it spends sending data.
A core function of LLS\_MIP is to schedule links to minimize the time spent on PAT.
This means that it will prioritize fewer, longer contacts over higher contact frequency.
This effectively increases the delay of data reaching the source while also increasing throughput.
This can also be measured as system uptime, where, in a 30-day experiment with 32 source nodes, LLS\_MIP had an uptime of just over 28 days, while algorithms that did not account for retargeting delay had an uptime of around 23 days.
While in practice, many other factors will affect link uptime, this is a substantial difference that significantly impacts the operational efficiency of an optical backhaul network.

\textit{(d) Algorithm Runtimes.} With any time-sliced dynamic topology, the execution time will increase exponentially with respect to the number of decision variables, following the product sum equation.
The tradeoff is that reducing the number of decision variables decreases the performance of the scheduled contact plan.
The model is capped at 2.5 hours of elapsed solve time, excluding time spent on pre-processing, building look-up tables, generating equations, or defining constraints.
Figure~\ref{fig:run_times} shows the impact on a log scale; as the network size increases linearly, the execution duration for LLS\_MIP increases exponentially.
LLS\_MIP reaches the timeout at just 16 nodes while LLS\_Greedy schedules the contact plan an order of magnitude faster, scaling linearly.

\textit{(e) Average Inter-Contact Delay.}
Figure~\ref{fig:source_node_delay} shows the average inter-contact delay, defined by the average time between source node transmission opportunities.
LLS\_Greedy is the only algorithm unable to maintain sub-30-minute scheduling intervals because it prioritizes capacity and fairness over delay.
LLS\_Greedy selects the nodes with the shortest pointing delays, thereby maximizing capacity.
This can lead to line-of-sight occlusions that restrict future contacts with a node and cause increased delay.
These delays correlate with the maximum buffer size and average buffer fullness during operation.
Even for delay-tolerant data, buffer management remains crucial due to mission SWaP-C constraints.

\textit{(f) Scheduled Fairness.}
We show fairness in Fig.~\ref{fig:scheduled_fairness} using Jain's fairness index, where the shared resource is the source node channel capacity measured as the amount of data transmitted by each source node.
Fairness is maximized if all source nodes are scheduled to transmit the same amount of data.
While LLS\_Greedy performs well for smaller networks, the greedy approach does not effectively load balance source node transmissions across the contact plan.
LLS\_MIP and FCP both have high fairness, where the MILP scheduler has less stable fairness that oscillates with the relay node count
This is because the MILP model employs a soft constraint for fairness, allowing it to fall within the defined bounds while improving capacity or creating feasible solutions.

\section{Conclusion}

FSO communications can dramatically increase the volume of scientific data returned from deep-space missions, but this performance depends on contact plans that respect the real pointing and acquisition delays of optical links. 
Our results show that accounting for these delays fundamentally alters the network's behavior. 
When retargeting delay is modeled accurately, the scheduler consistently prefers fewer but longer contacts, thereby reducing retargeting overhead and increasing the fraction of each window spent transmitting data. 
This shift yields more than a 30\% increase in scheduled backhaul capacity compared to state-of-the-art greedy models under realistic orbital dynamics, demonstrating an improvement significant enough to influence how future missions plan their downlinks. 
The analysis also shows that ignoring retargeting delay leads to schedules that appear efficient in theory but are infeasible or severely overoptimistic in practice. 
These findings highlight the importance of PAT-aware temporal modeling for any autonomous optical backhaul system that aims to operate reliably at interplanetary scales. 
Future work will extend this approach to evaluate end-to-end routing performance over the resulting retargeting-aware contact plans. 


\bibliographystyle{IEEEtaes}
\bibliography{references}

\end{document}